\documentclass{iau}
\usepackage{graphicx}

\def\d3k{{\displaystyle {\rm d}{\bf k} \over \displaystyle (2\pi)^3}}

\renewcommand\vec[1]{\boldsymbol{#1}}
\newcommand{\grad}{\vec{\nabla}}
\newcommand{\D}{D_{+}}

\title[The Zeldovich \& Adhesion approximations] 
{The Zeldovich \& Adhesion approximations\\{\Large and applications to the local universe}}

\author[Johan Hidding]   
{Johan Hidding$^1$
\and Rien van de Weygaert$^1$
\and Sergei Shandarin$^2$}

\affiliation{$^1$
Kapteyn Astronomical Institute, Univ. Groningen, P.O. Box 800, Groningen, the Netherlands \\ email: {\tt johannes.hidding@gmail.com} \\[\affilskip]
$^2$Department of Physics and Astronomy, Univ. Kansas, Lawrence, KS 66045-7582}

\pubyear{2014}
\volume{308}  
\pagerange{119--126}
\jname{The Zeldovich Universe: Genesis and Growth of the Cosmic Web}
\editors{R. van de Weygaert, S.F. Shandarin, E. Saar \& J. Einasto, eds.}

\begin{document}

\maketitle

\begin{abstract}
The Zeldovich approximation (ZA) predicts the formation of a web of
singularities.  While these singularities may only exist in the most formal
interpretation of the ZA, they provide a powerful tool for the analysis of
initial conditions. We present a novel method to find the skeleton of the
resulting cosmic web based on singularities in the primordial deformation
tensor and its higher order derivatives.  We show that the $A_3$-lines predict
the formation of filaments in a two-dimensional model. We continue with
applications of the adhesion model to visualise structures in the local ($z <
0.03$) universe.
\end{abstract}
\firstsection 
\section{The Zeldovich approximation}
The Zeldovich Approximation (ZA) (\cite{Zeldovich1970,Shandarin1989}) describes 
structure formation in the form of a deceptively simple equation
\begin{equation}
\vec{x}(\vec{q}, t) = \vec{q} - \D(t)\grad\Phi_0(\vec{q}).
\label{eq:zeld}
\end{equation}
Rather than describing just ballistic motion, this equation hides a formalism of
Lagrangian collision-free fluid mechanics. In the context of emerging interest
in phase-space folding descriptions of structure formation 
(\cite{Shandarin2012,Falck2012,Abel2012}), 
it becomes all the more relevant to understand this expression at a much deeper
level. We can see why there is more to the ZA than inertial motion, if we compute
particle densities from the above expression.  Density increases or decreases
locally as a fluid element contracts or expands.  Taking a fluid element from
Lagrangian location $\vec{q}$, we can quantify its deformation in terms of the
deformation tensor $d_{ij} = - \partial_i \partial_j \Phi_0$.  This tensor is
best studied locally in the eigenvector frame of reference
$\{\vec{e}_{\lambda}\}$, where $d_{ij}$ becomes diagonal.  The density is then 
\begin{equation}
\delta(\vec{x}) + 1 
	= \sum_{q \in \{q^{\star}\}} \bigg|\det \frac{\partial x_i}{\partial q_j}\bigg|_q^{-1} 
	= \sum_{q \in \{q^{\star}\}} \bigg|\prod_i \big(1 - \D \lambda_i\big)\bigg|_q^{-1},
\end{equation}
where $\{\vec{q}^{\star}\}$ is the set of Lagrangian locations solving for
$\vec{x}$ in equation \ref{eq:zeld}, and $\lambda_i$ are the eigenvalues of the
deformation tensor $d_{ij}$.  Note that this expression for the density has
singularities whenever for one of the eigenvalues we have 
\begin{equation}
\lambda_i = 1/\D(t).
\label{eq:a2}
\end{equation}

\begin{figure}[t]
\includegraphics[width=\textwidth]{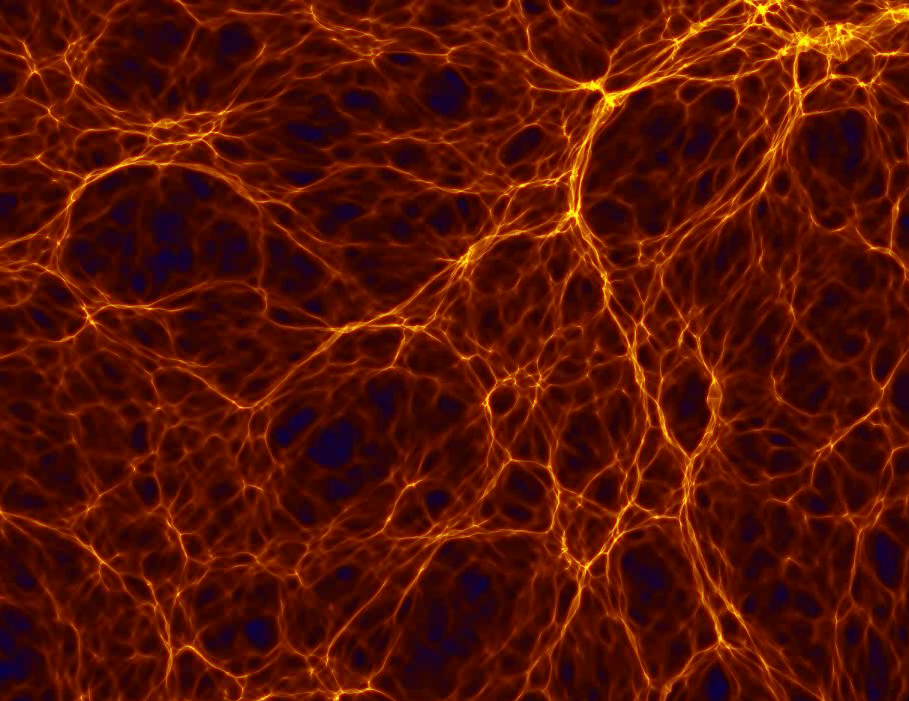}
\caption{The Zeldovich approximation. Density map of the cosmic structure following the 
evolution according to the Zeldovich formalism. The cosmic web is sharply rendered, 
with most of the structures residing just before or around shell crossing. From Hidding 2016.}
\label{fig:zeldovich}
\end{figure}

\noindent This notion gives us the traditional interpretation of ZA, namely one where
gravitational collapse occurs in three possible stages. First objects form
along the major axis of collapse, making pancakes; then along the second
eigenvector filaments form; and finally if and when all three eigenvalues have
passed a singularity, a cluster forms. This corresponds to the \emph{Morse
theory} view of nodes and saddles exploited in many structure finders. 
We show that, even in the narrow confines of the ZA, this interpretation is not
complete; even that it is wrong on the account of the formation of the first
filaments (\cite{Hidding2014}).  Taking the mathematics of
Morse theory a step further, we arrive at Lagrangian \emph{singularity theory}
(\cite{Arnold1982,Arnold1986}).  This theory
shows how to predict the evolution of \emph{folds}, \emph{cusps},
\emph{swallow-tails}, \emph{butterflies} and \emph{umbilics} directly from the
initial velocity potential $\Phi_0$. Due to the relative complexity of this
method we are forced to restrict our further discussion to the two-dimensional
case.\\

\begin{figure}[t]
\includegraphics[width=\textwidth]{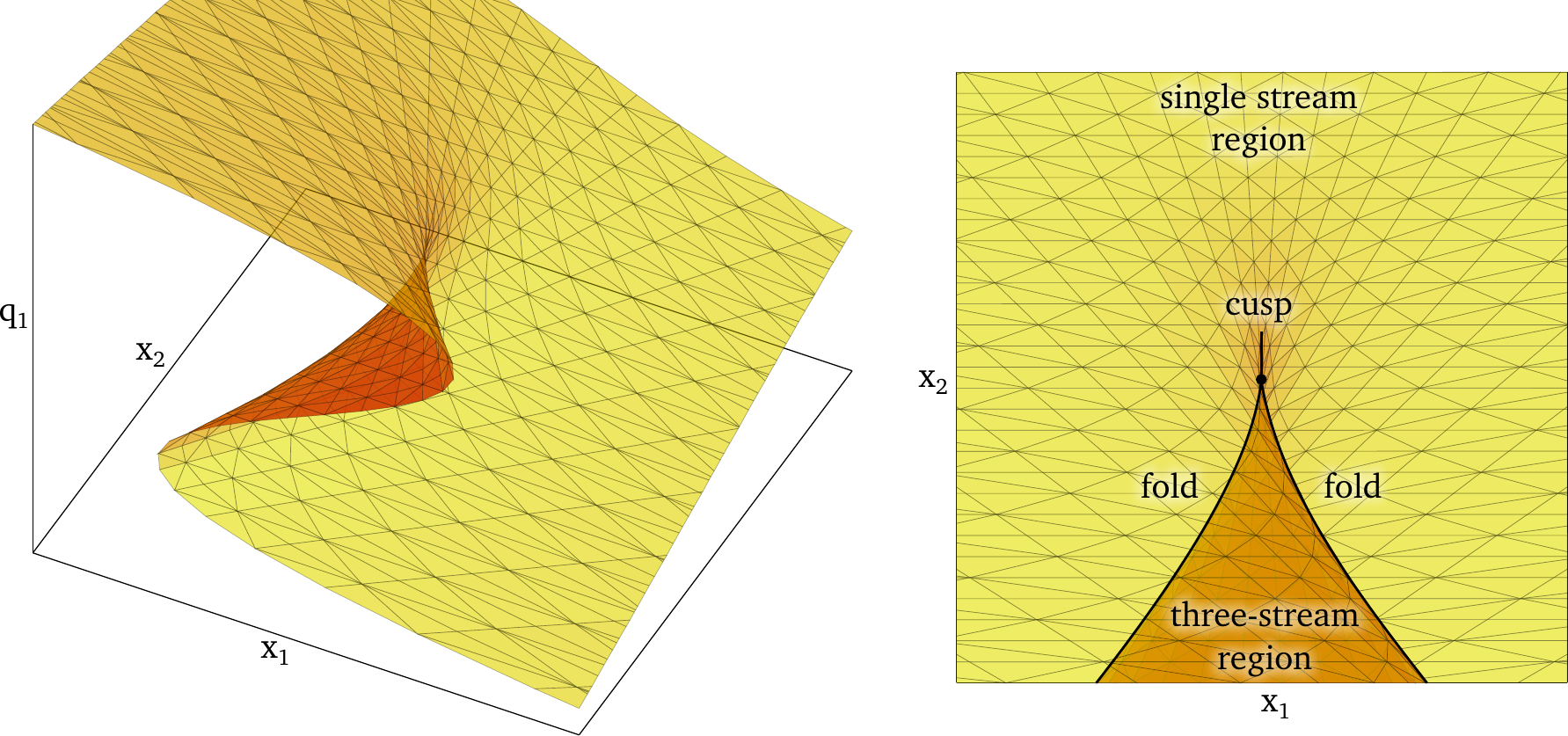}
\caption{{\it The cusp singularity.} Singularities arise if we project a smooth
manifold down one or more directions. In this case we show one Lagrangian
direction as the $z$-axis and the two remaining axes in Eulerian space. On the
right is the projected view, with the fold and cusp locations marked.  Between
the folds we find a three-stream region.}
\label{fig:cusp}
\end{figure}

{\underline{\it Formation of pancakes.}} A \emph{fold} is the simplest kind of
singularity we have. It is the caustic that separates single-stream from multi-stream
regions and is also found under the cryptic name $A_2$ 
\footnote{$A_2$
refers to the ADE classification of singularities introduced by Arnold. In this
paper we also deal with $A_3$ for cusp, $A_4$ for swallow-tail, and $D_4$ for
umbilic singularities.}.
At any moment in time, pancakes can be identified as the locations of $A_2$
folds. At a fold the phase-space sheet (see Fig. \ref{fig:cusp}) is tangent to
line of projection. In the case of ZA, this happens when $\lambda = 1/\D$,
identifying the level-sets of $\lambda$ as the Lagrangian progenitors of folds.
Two folds may connect at a \emph{cusp}. A fold being a line of tangency on the
phase-space sheet, there exists points where the fold line itself is tangent
to the projection, these points are the cusps (see Fig. \ref{fig:cusp}).  In
the tensor field $d_{ij}$, a cusp is found where a level-set of $\lambda$ is
tangent to the corresponding eigenvector $\vec{e}_{\lambda}$, or
\begin{equation}
\grad \lambda \cdot \vec{e}_{\lambda} = 0.
\end{equation}
Finding all $A_3$-points for each level-set of $\lambda$ traces an $A_3$-line.
The set of $A_3$-lines trace the entire network of filaments formed in the ZA,
throughout time. It can be shown that all maxima and saddle points of the
function $\lambda(\vec{q})$ also lie on an $A_3$-line. Lowering a
level-plane down on the function $\lambda(\vec{q})$, we can see that at the
maxima of $\lambda$ ($A_3^+$-points) two cusps are created, while at the saddle
points ($A_3^-$-points) they annihilate, merging two pancakes.  $A_3$-lines
terminate only in $D_4$ \emph{umbilic} points, but we choose to also truncate
them where $\lambda = 0$, beyond which they loose their physical
significance.\\

\begin{figure}[b]
\includegraphics[width=\textwidth]{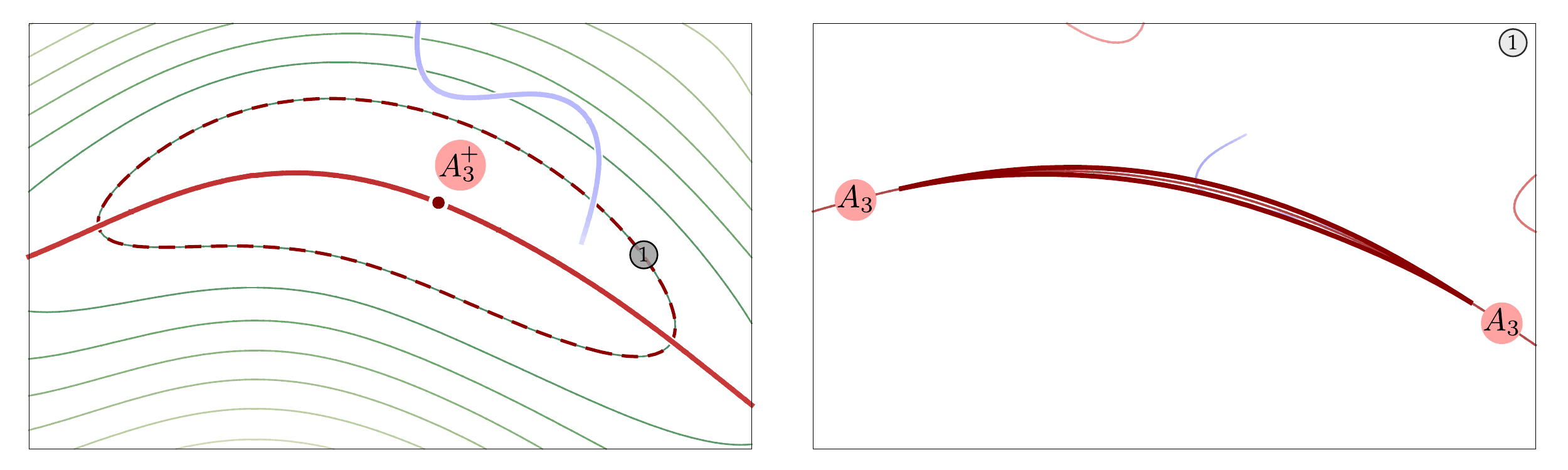}
\caption{{\it Pancake genesis.} On the left we see in Lagrangian space the
contours of the first eigenvalue around an $A_3^+$-point.  The
Eulerian counterpart of the dashed contour is shown on the right. It shows
the pancake at its prime, highly elongated with a cusp on each end.}
\label{fig:pancake}
\end{figure}


\begin{figure}[t!]
\includegraphics[width=\textwidth]{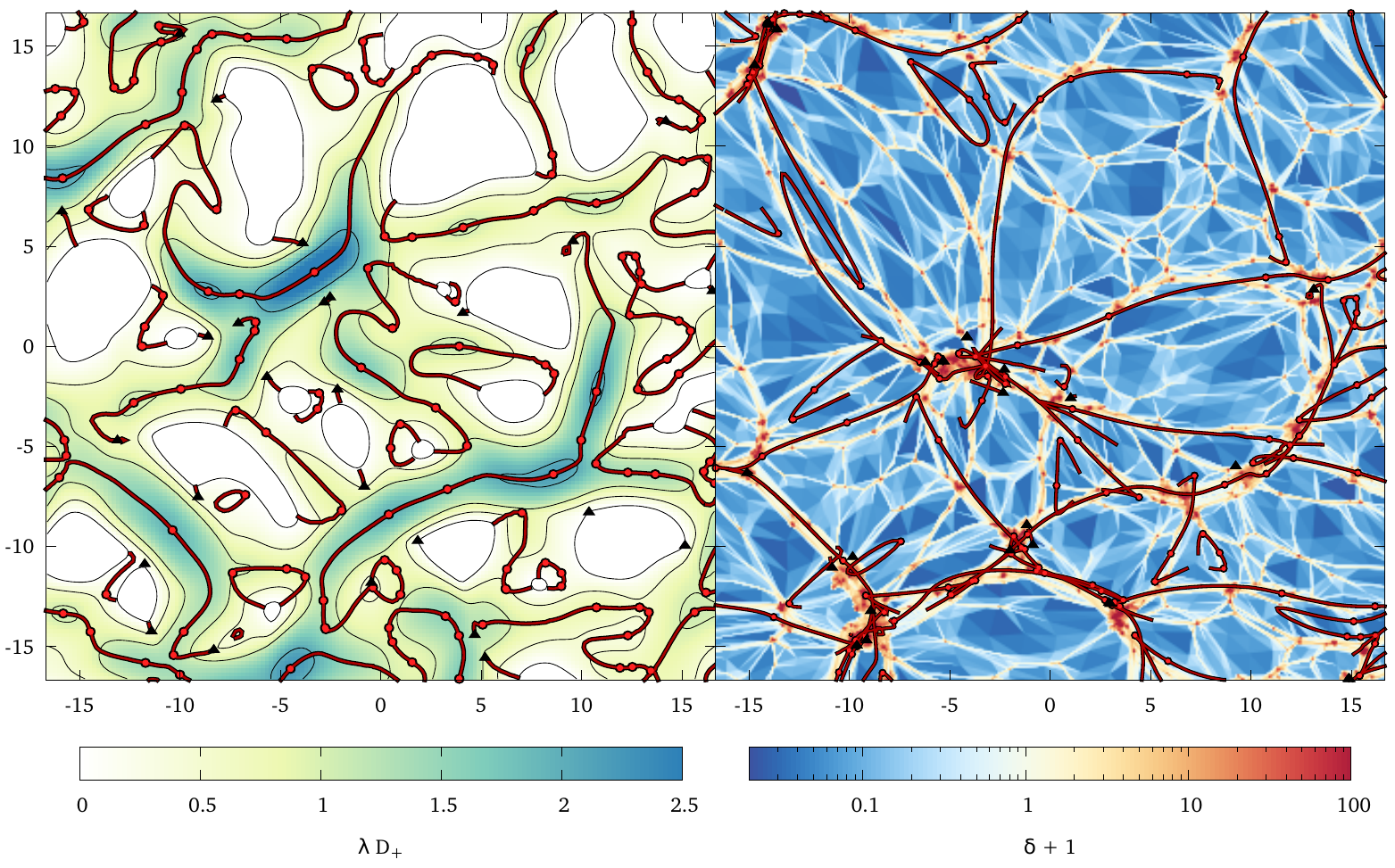}
\caption{{\it Comparison with N-body.} On the left: ZA on filtered initial
conditions; contours show the first eigenvalue, $A_3$-lines in red, and
triangles showing $D_4$ points. On the right: the density contrast resulting
from a 2D PM code, with the Eulerian displaced $A_3$-lines over-plotted. }
\label{fig:nbody}
\end{figure}

{\underline{\it Splitting of pancakes.}} A pancake may branch by creating two new cusps
at a fold. Typically one of these cusps remains within the confines of the
present pancake and the other dashes out to create a subsidiary pancake, often
to merge later with another pancake at an $A_3^-$-point, creating a
three-legged structure. The point at which a pancake branches is called a
swallow-tail, denoted $A_4$.  An $A_4$ singularity is found in
Lagrangian space at points where an $A_3$-line is tangent to the corresponding
eigenvector, or equivalently at local maxima of $\lambda$ limited to the
$A_3$-line. Important to note here, is that we don't need to involve the second
eigenvalue to create a node in the network of caustics. Moving this discussion
to the three-dimensional case, we don't strictly need to collapse along the
second eigenvector to create a filament-like structure. This also suggests the
existence of different possible late-time morphologies for filaments.\\


{\underline{\it Scaling and comparison with N-body.}} We computed the
$A_3$-lines for a set of initial conditions and compared the result with those
of a 2D N-body code. A non-linear time evolution may be approximated by
truncating power of the initial conditions at scales smaller than the scale of
non-linearity. It is well known that observable filaments have a density
contrast around unity, so this method should give realistic results. We find
good agreement for all relevant $P(k) \propto k^n$ power-spectra, in an
eye-ball comparison of filaments predicted by ZA with density fields from a 2D
PM code. An example is given in Fig. \ref{fig:nbody}.

\section{The Adhesion approximation}
\begin{figure}[t]
\includegraphics[width=\textwidth]{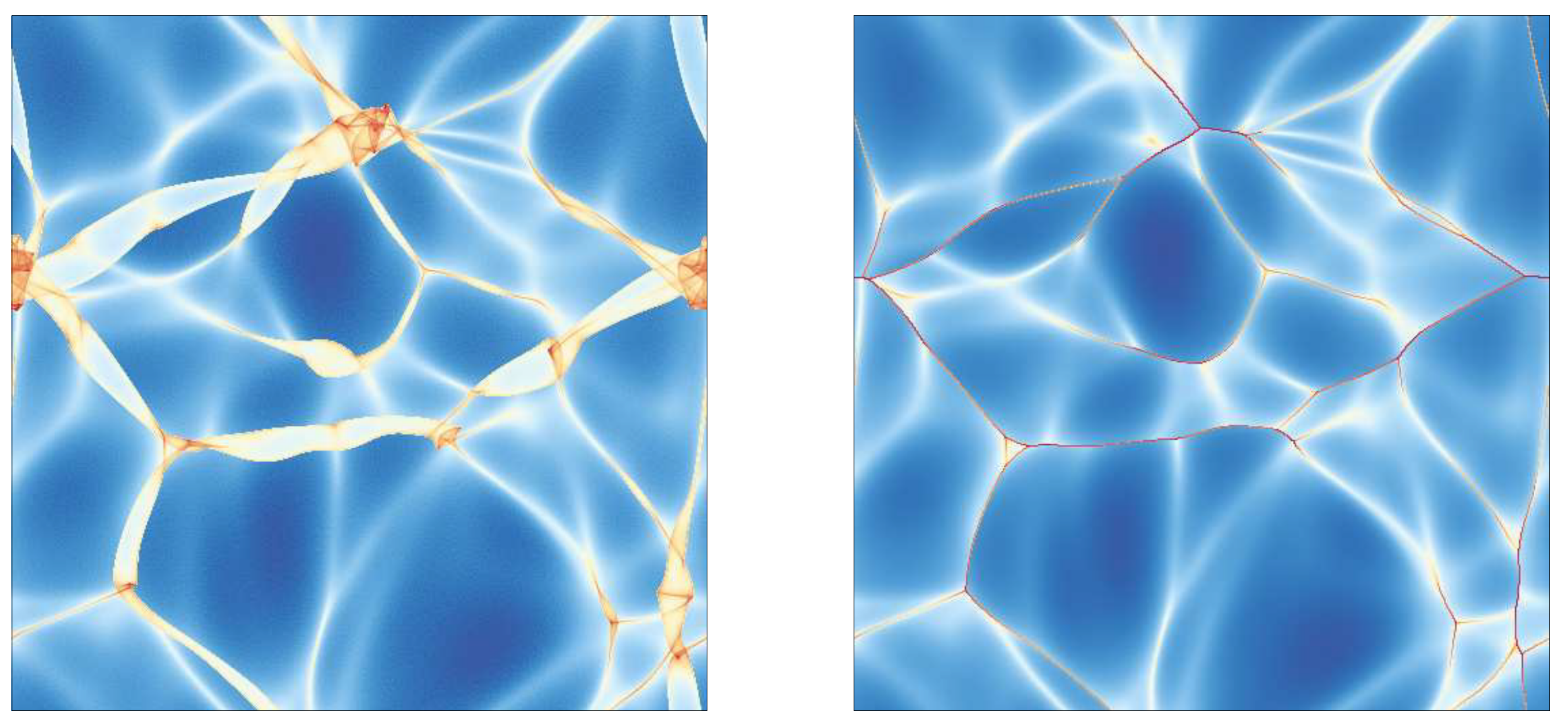}
\caption{{\it ZA and adhesion compared.} Outside multi-stream regions the
results from the ZA and adhesion are identical.  Adhesion contains an
artificial viscosity term that only `activates' when streams cross.
Multi-stream regions are thus collapsed to infinitesimally thin structures.}
\label{fig:adhesion}
\end{figure}

\begin{figure}[b]
\centering
\includegraphics[width=0.49\textwidth]{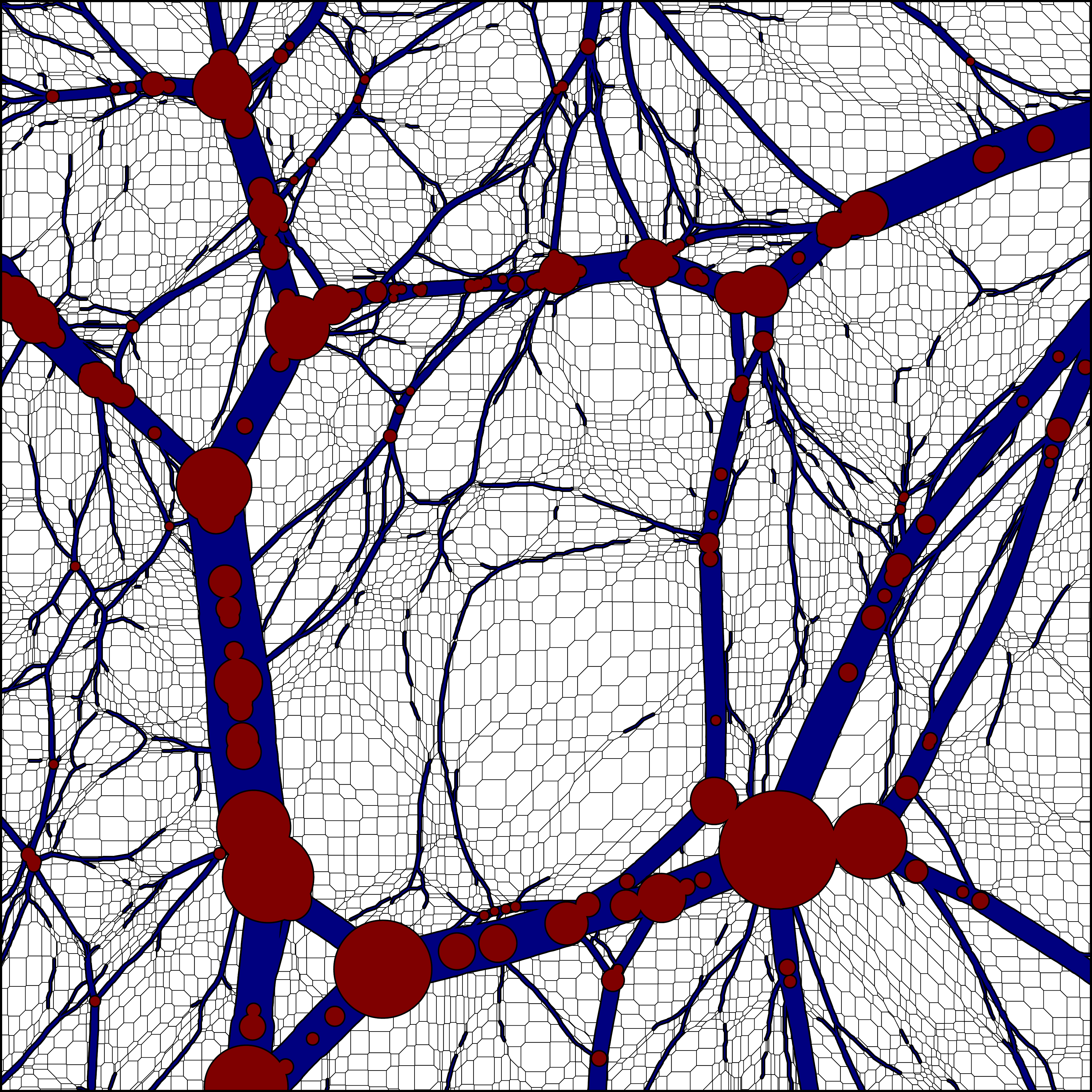}
\includegraphics[width=0.49\textwidth]{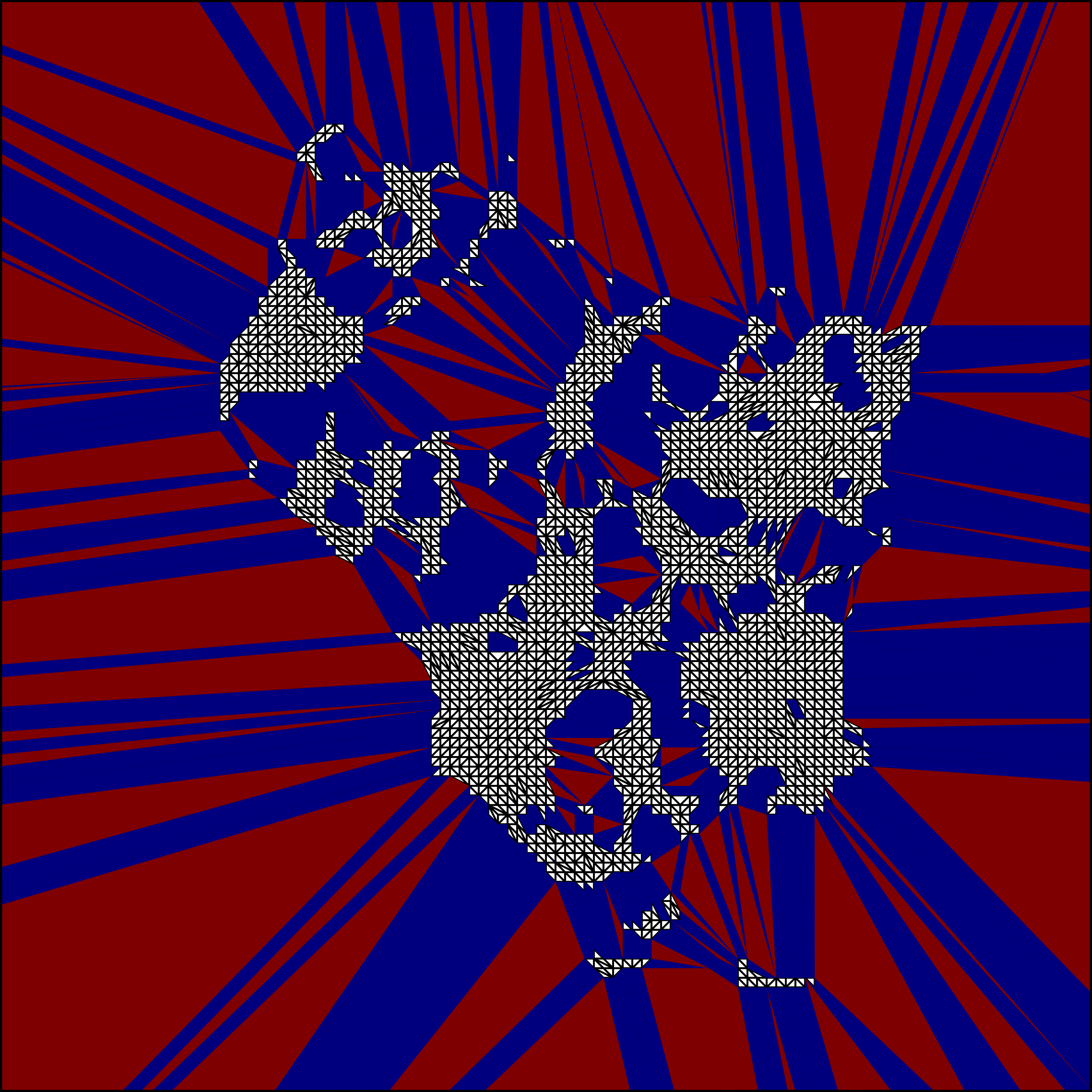}
\caption{{\it Dual structures in adhesion.}}
\label{fig:dual}
\end{figure}

We showed how the emergence of caustics in the ZA allows us to trace the
formation of cosmic structures in a formal, yet physically meaningful way.
However, the ZA suffers from a major flaw in that it doesn't allow for
gravitational interaction, and therefore hierarchical structure formation.
This is because the ZA is solely based on local analysis of the velocity
potential and its chain of derivatives. The adhesion model moves beyond local
considerations, which makes it computationally more intensive than the ZA (though
still much faster than N-body). Still, results are computed from
initial conditions directly and with complete accountability.
The adhesion model is arrived at by taking the source-free (hence collision-free)
Euler equation, and adding an artificial viscosity term to \emph{emulate}
the effects of gravity (\cite{Gurbatov1984,Shandarin1989}). The resulting equation
\begin{equation}
\partial_t \vec{u} + (\vec{u} \cdot \grad) \vec{u} = \nu \nabla^2 \vec{u},
\end{equation}
is known as Burgers' equation; in the limit where $\nu \to 0$, it has the exact solution
\begin{equation}
\Phi(\vec{x}, t) = \max_q \left(\Phi_0(\vec{q}) - \frac{(\vec{x} - \vec{q})^2}{2 \D(t)}\right).
\label{eq:burgsol}
\end{equation}
The \emph{global maximum} in this solution can be computed efficiently using
either a Legendre transform, convex hull (\cite{Vergassola1994}) or a weighted
Voronoi diagram (\cite{Hidding2012,Hidding2016a}). One condition for reaching
an extremum is that the first derivative of the maximised quantity should
vanish.  Performing this test reduces above equation to the ZA as presented in
equation \ref{eq:zeld}.  The global maximisation guarantees that the resulting
map from Lagrangian to Eulerian space stays \emph{monotonic} always.
Where and whenever shell-crossing occurs in the ZA, adhesion creates a solid
structure. Matter inside these structures is confined to stay inside, which is
the reason people may refer to the adhesion model as having ``sticky particles''.
Outside collapsed structures the results from the ZA
and adhesion are identical; caustics from ZA are compressed to infinitesimally
thin structures (see Fig. \ref{fig:adhesion}). This unifies Zeldovich' idea of 
collapsed structures in terms of shell crossing with a hierarchical formation model.\\

\begin{figure}[b]
\mbox{\hskip -0.45cm\includegraphics[width=\textwidth]{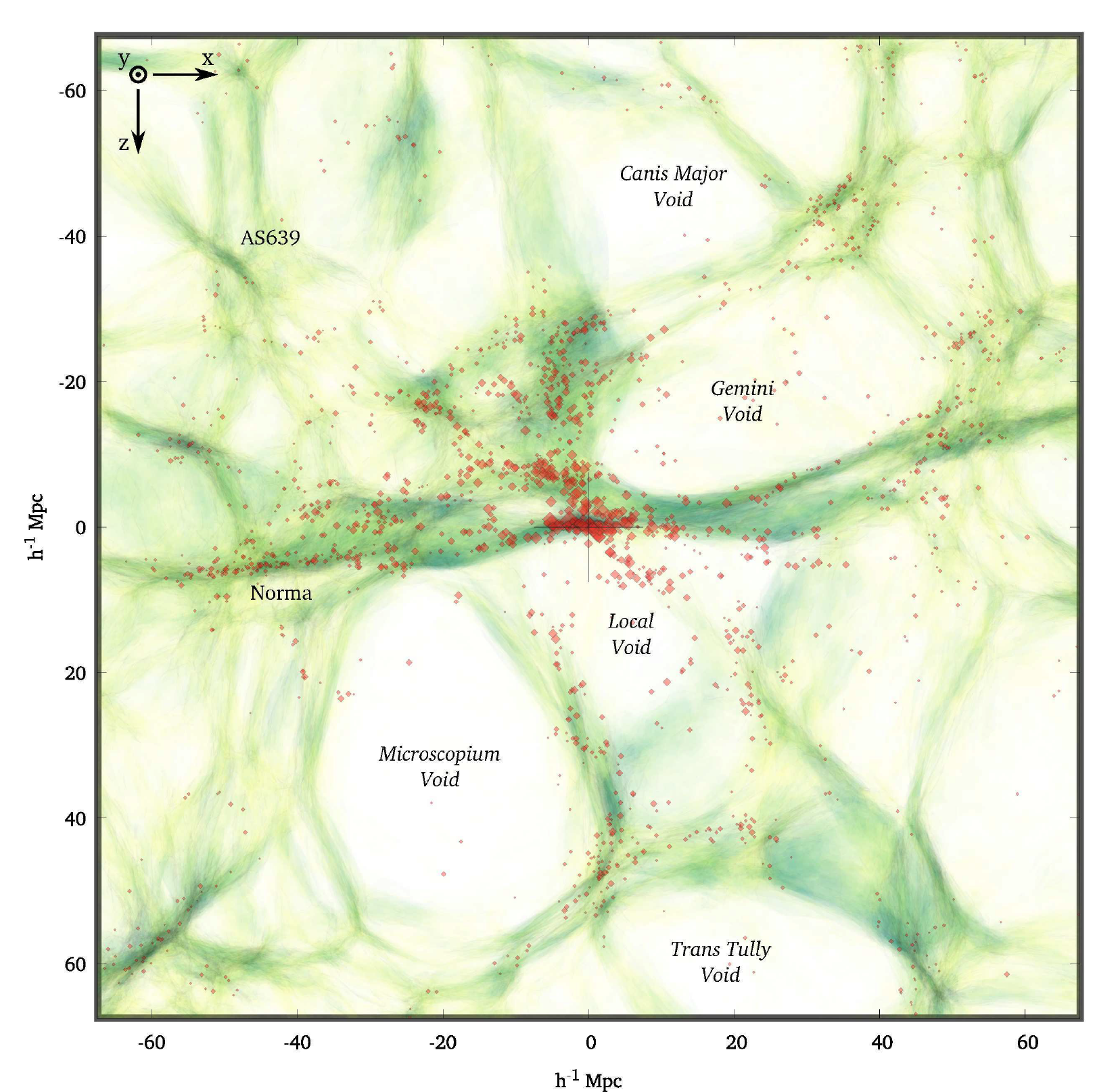}}
\caption{{\it Local Universe adhesion reconstruction.} Structure of the lcoal Universe. 
The reconstruction of the weblike structure in the local Universe, 
sampled by the 2MRS survey, has been obtained on the basis of the adhesion formalism applied to a set 
of 25 constrained Bayesian KIGEN realizations of the primordial density and velocity field in the 
Local Universe. The image shows the density field in a 10 Mpc thick slice perpendicular to the 
plane of the Local Supercluster. Note that the density field concerns the dark matter distribution. The red dots 
are the 2MRS galaxies in the same volume. From Hidding 2016 and Hidding et al. 2016b, using initial 
conditions from He{\ss} et al. 2013.}
\label{fig:localvoids}
\end{figure}

{\underline{\it Dual geometry.}} The solution to Burgers' equation given in
expression \ref{eq:burgsol} is identical to the definition of the \emph{weighted
Voronoi tessellation}, weighted by the potential. The Voronoi cell of a Lagrangian point
$\vec{q} \in \mathcal{L}$ occupies an area in Eulerian space $\mathcal{E}$ given by
\begin{equation}
V_q = \left\{\vec{x} \in \mathcal{E}\ \Big|\ (\vec{x} - \vec{q})^2 + w_q\leq
(\vec{x} - \vec{p})^2 + w_p,\ \forall\ \vec{p} \in \mathcal{L}\right\}.
\end{equation}
Taking $w(\vec{q}) = 2 \D \Phi_0(\vec{q})$ as the weights in this expression
reduces it to the given solution in equation \ref{eq:burgsol}. Where there is a
Voronoi tessellation, there is its dual: the \emph{Delaunay triangulation}.
It is the latter that gives us the origin and mass of matter residing in the nodes,
edges and faces of the Voronoi tessellation (see Fig. \ref{fig:dual}).

\section{The Local Universe}
One application of the adhesion model is the detection of walls, filaments and nodes
in cases where some form of an initial potential is available. We ran our adhesion code
on a set of 25 constrained initial conditions \emph{reconstructed} (\cite{Kitaura2013,Hess2013}) to produce structures in our
local universe ($z < 0.03$) (\cite{Hidding2016,Hidding2016b}). This reconstruction is based on the 
2MASS redshift catalog (\cite{Huchra2012}), which covers the full sky except for galactic lattitudes $|b| < 5^{\circ}$. 

\begin{figure}[t]
\includegraphics[width=\textwidth]{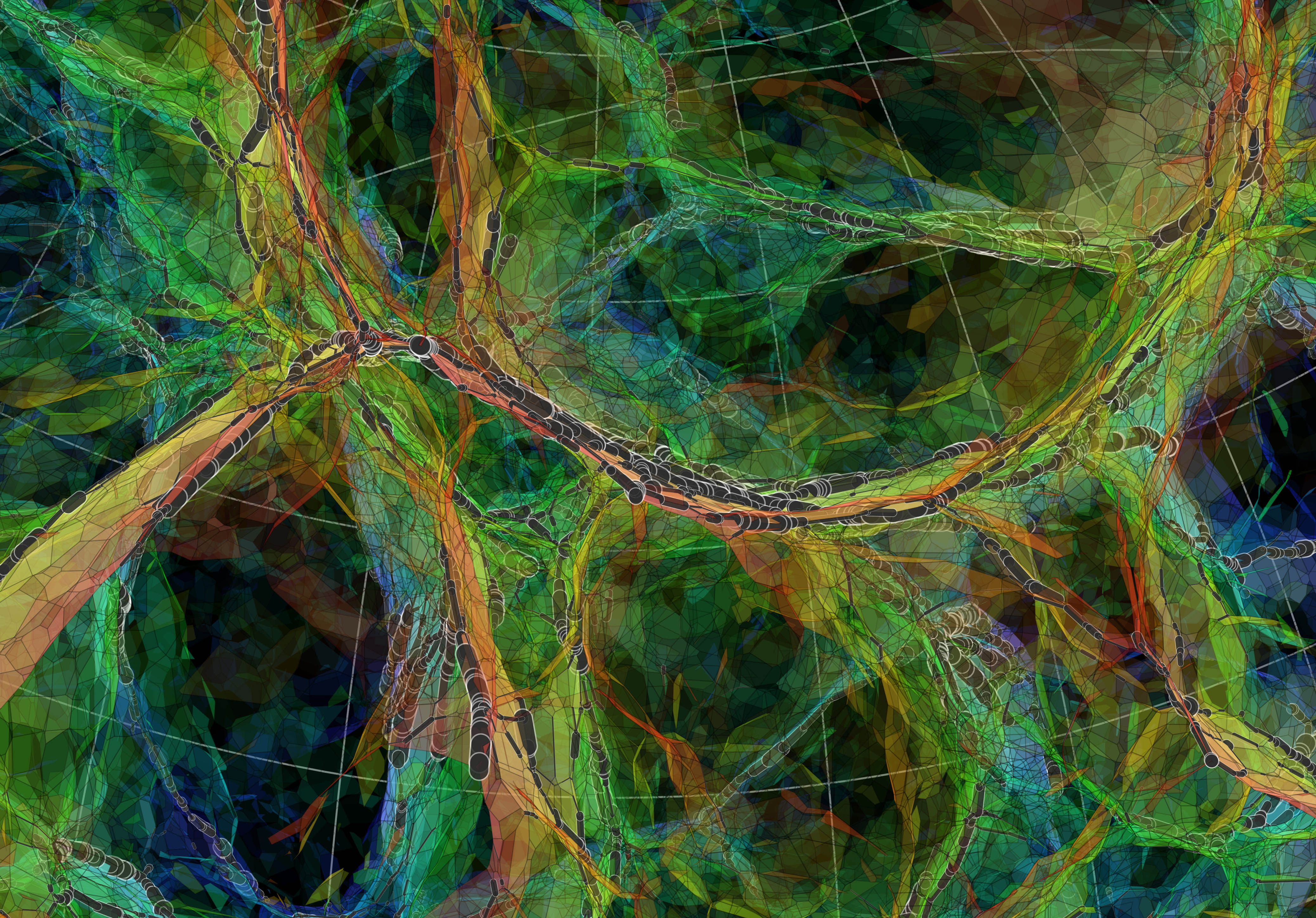}
\caption{{\it The Pisces-Perseus Supercluster.} A 3-D isodensity surface rendering of the 
intricate filamentary structure around the Pisces-Perseus supercluster. It is based on the 
adhesion reconstruction of the local Cosmic Web, based on constrained realizations of the 
local primordial density and velocity field implied by the 2MRS galaxy redshift survey. 
From Hidding 2016 and Hidding et al. 2016b, using initial conditions from He{\ss} et al. 2013.}
\label{fig:perseus}
\end{figure}

Figure~\ref{fig:localvoids} provides a remarkably detailed reconstruction of the cosmic web in the 2MRS volume. It 
shows the (surface) density of the weblike structures in the Local Universe. These are the result of adhesion 
simulations by \cite{Hidding2016} and \cite{Hidding2016b}, based on the the constrained Bayesian KIGEN reconstruction by \cite{Kitaura2013} 
of the initial conditions in the local volume traced by the 2MRS redshift survey. For a given 
Gaussian primordial field, the adhesion formalism allows the accurate reconstruction of the rich pattern 
of weblike features that emerge in the same region as a result of gravitational evolution. The adhesion formalism 
was applied to 25 constrained realizations of the 2MRS based primordial density field (\cite{Hidding2012,Hidding2016}). 
The mean of these realizations gives a reasonably accurate representation of the significant filamentary and wall-like 
features in the Local Universe. Most outstanding is the clear outline of the void population 
in the local Universe. The reconstruction also includes the velocity flow in the same cosmic region. It reveals the 
prominent nature of the outflow from the underdense voids, clearly forming a key aspect of the dynamics of the 
Megaparsec scale universe. 

The Local Universe structure in figure~\ref{fig:localvoids} presents a telling image of a void dominated 
large scale Universe. Many of the voids in the adhesion reconstruction can be identified with the void 
nomenclature proposed by Fairall (\cite{fairall1998}), who mainly identified these voids by eye from the 
6dFGRS survey. It is interesting to see that the socalled Tully void appears to be a richly structured underdense 
region, containing at least the Microscopium Void, the Local Void and the "Trans Tully Void". 

In the same reconstruction, we are studying the intricate filamentary network in and around the 
Pisces-Perseus supercluster. The image in figure\ref{fig:perseus} provides a nice impresssion of the 
complex 3-dimensional structure and connectivity along the main ridge of the Pisces-Perseus supercluster. 
It also shows how the main ridge connects to several neighbouring filaments, connecting near massive 
clusters along the ridge, and how these filaments surround a lower density planar structure. Interesting 
is to note the clustering and alignment of the small (whitish) filamentary tendrils in and around the 
main arteries of the Cosmic Web. Analysis 
of this weblike structures region is under progress and will be first reported in \cite{Hidding2016} and 
\cite{Hidding2016b}.
\acknowledgements 
We wish to acknowledge Francisco Kitaura and Steffen He{\ss} for the collaboration 
on the Local Universe reconstruction project, and for allowing the use of figure 7 and 8 
in this report in advance of publication in \cite{Hidding2016b}.


\end{document}